\documentclass[submission, Phys]{SciPost}

\pdfoutput=1

\usepackage[utf8]{inputenc} %
\usepackage[T1]{fontenc} 	%
\usepackage[english]{babel} %

\usepackage[bitstream-charter]{mathdesign}
\usepackage{geometry} 		%
\usepackage{amsmath} 		%
\usepackage{mathtools} 		%
\usepackage{float} 			%
\usepackage{graphicx} 		%
\usepackage{tabularx} 		%
\usepackage{booktabs} 		%
\usepackage{color, xcolor} 	%
\usepackage{commath}
\usepackage{pdfpages} 		%
\usepackage{extarrows} 		%
\usepackage{multirow} 		%
\usepackage{multicol} 		%
\usepackage{caption} 		%
\usepackage{subcaption} 	%
\usepackage{enumitem} 		%
\usepackage{xspace} 		%
\usepackage{stackrel} 		%
\usepackage{tikz} 			%
\usetikzlibrary{calc}
\usepackage{braket} 		%
\usepackage{bm} 			%
\usepackage{tensor} 		%
\usepackage{slashed} 		%
\usepackage{siunitx} %
\usepackage{lastpage} 		%
\usepackage{cite} 			%
\usepackage[normalem]{ulem} %
\usepackage{fontawesome} 	%
\usepackage{tocloft} 		%
\usepackage{titlesec} 		%
\usepackage{doi} 			%
\usepackage[capitalize]{cleveref} 	%
\usepackage[nottoc, notlot, notlof]{tocbibind} 	%
\usepackage[ruled, vlined]{algorithm2e} 		%
\usepackage{makecell}
\usepackage{pifont}
\usepackage{stackrel}

\hypersetup{
	pdftitle={Returning CP-Observables to The Frames They Belong},
	pdfauthor={Jona Ackerschott, Rahool Kumar Barman, Dorival Gonçalves,
        Theo Heimel, and Tilman Plehn},
    linktoc=section,
	breaklinks=true,			%
	colorlinks=true, 			%
	linkcolor={red!50!black}, 	%
	citecolor={blue!50!black}, 	%
	urlcolor={blue!80!black} 	%
} 

\DeclareSymbolFont{usualmathcal}{OMS}{cmsy}{m}{n}
\DeclareSymbolFontAlphabet{\mathcal}{usualmathcal}

\SetArgSty{textnormal}
\SetKwComment{Comment}{{\small\#}~}{}
\SetCommentSty{mycommfont}

\setitemize{itemsep=2pt, parsep=0pt} 				%
\setenumerate{itemsep=2pt, parsep=0pt} 				%
\setitemize{itemsep=2pt,topsep=2pt,parsep=0pt,partopsep=0pt,leftmargin=*}
\setenumerate{itemsep=0pt,topsep=2pt,parsep=0pt,partopsep=0pt,labelindent=3pt,leftmargin=*}
\newcommand{\Langle}{\bigl\langle}
\newcommand{\Rangle}{\bigr\rangle}

\newcommand{\qqquad}{\qquad\quad}

\newcommand{\sign}{\operatorname{sign}} 	%

\newcommand{\lag}{\mathscr{L}}
\newcommand{\ope}{\mathcal{O}}
\newcommand{\obs}{\mathcal{O}}
\newcommand{\loss}{\mathcal{L}} 	%

\newcommand{\pytorch}{\textsc{PyTorch}\xspace}

\newcommand{\adam}{\textsc{Adam}\xspace}

\newcommand{\arXiv}[2][]{%
	\ifthenelse{\equal{#1}{}}%
	{\href{http://arxiv.org/abs/#2}{arXiv:#2}}%
	{\href{http://arxiv.org/abs/#2}{arXiv:#2~[#1]}}}

\newcommand{\gev}{\text{GeV}}

\def\slashchar#1{\setbox0=\hbox{$#1$}           %
   \dimen0=\wd0                                 %
   \setbox1=\hbox{/} \dimen1=\wd1               %
   \ifdim\dimen0>\dimen1                        %
      \rlap{\hbox to \dimen0{\hfil/\hfil}}      %
      #1                                        %
   \else                                        %
      \rlap{\hbox to \dimen1{\hfil$#1$\hfil}}   %
      /                                         %
   \fi}

\newcommand{\tikznode}[2]{%
\ifmmode%
\tikz[remember picture,baseline=(#1.base),inner sep=0pt] \node (#1) {$#2$};%
\else
\tikz[remember picture,baseline=(#1.base),inner sep=0pt] \node (#1) {#2};%
\fi}

\def\mathswitchr#1{\relax\ifmmode{\text{#1}}\else$\text{#1}$\xspace\fi}
\def\mathswitch#1{\relax\ifmmode#1\else$#1$\xspace\fi}

\graphicspath{{./figures/}}

\begin{document}

\begin{center} 
    {\Large\textbf{Returning CP-Observables to The Frames They Belong}}
\end{center}

\begin{center}
  Jona Ackerschott\textsuperscript{1},
  Rahool Kumar Barman\textsuperscript{2},
  Dorival Gon\c{c}alves\textsuperscript{2}, \\
  Theo Heimel\textsuperscript{1}, and 
  Tilman Plehn\textsuperscript{1}
\end{center}

\begin{center}
    {\bf 1} Institut f\"{u}r Theoretische Physik, Universit\"{a}t Heidelberg, Germany\\
    {\bf 2} Department of Physics, Oklahoma State University, Stillwater, USA
\end{center}

\begin{center}
    \today
\end{center}
 
\section*{Abstract}
         {\bf Optimal kinematic observables are often defined in
           specific frames and then approximated at the reconstruction
           level. We show how multi-dimensional unfolding methods
           allow us to reconstruct these observables in their proper
           rest frame and in a probabilistically faithful way. We
           illustrate our approach with a measurement of a
           CP-phase in the top Yukawa
           coupling. Our method makes use of key advantages of
           generative unfolding, but as a constructed observable it
           fits into standard LHC analysis frameworks.}

\vspace{10pt}
\noindent\rule{\textwidth}{1pt}
\tableofcontents\thispagestyle{fancy}
\noindent\rule{\textwidth}{1pt}
\vspace{10pt}

\clearpage
\section{Introduction}
\label{sec:intro}

With the LHC continuing its success story of precision hadron collider
physics, the size and complexity of the datasets of the upcoming Run~3
and HL-LHC are challenging the existing analysis
methodology~\cite{Campbell:2022qmc,Butter:2022rso,Plehn:2022ftl}. At
the same time, the goal of LHC physics has moved from model-based
searches for physics beyond the Standard Model (SM) to a comprehensive
analysis of all its data, based on consistent analysis frameworks like
the Standard Model effective
theory~\cite{Buchmuller:1985jz,Grzadkowski:2010es,Brivio:2017vri}.

The first step in any global analysis based on the fundamental
principles of QFT is to determine the underlying symmetries, which are
required to construct the effective Lagrangian. The, arguably, most
interesting symmetry in the SM is $CP$, linked to cosmology through
the Sakharov conditions for baryogenesis~\cite{Sakharov:1967dj}, and
potentially realized in an extended Higgs
sector~\cite{Basler:2021kgq}. In the language of effective theory,
$CP$-violation in the Higgs coupling to vector bosons is
loop-suppressed and arises at dimension
six~\cite{Plehn:2001nj,Englert:2012xt,Dolan:2014upa,Brehmer:2017lrt,Bernlochner:2018opw}.
In contrast, $CP$-violation in Higgs couplings to fermions can appear
at dimension four~\cite{Buckley:2015vsa}, making a $CP$-phase in the
top Yukawa coupling the most sensitive link between baryogenesis and
LHC
physics~\cite{Ellis:2013yxa,Boudjema:2015nda,Buckley:2015ctj,Gritsan:2016hjl,Goncalves:2016qhh,
Mileo:2016mxg,AmorDosSantos:2017ayi,Azevedo:2017qiz,Goncalves:2018agy,ATLAS:2018mme,
CMS:2018uxb,Ren:2019xhp,Martini:2021uey,Goncalves:2021dcu,Barman:2021yfh,Bahl:2021dnc,
Barman:2022pip,Azevedo:2022jnd}.

Obviously, we do not want to leave the test of fundamental Lagrangian
symmetries to a global analysis~\cite{Englert:2019xhk,Bahl:2022yrs}
with limited control over experimental and theoretical
uncertainties~\cite{Biekotter:2018ohn,Brivio:2019ius,Ellis:2020unq,Brivio:2021alv,
Ethier:2021bye},
including systematics from parton
densities~\cite{Greljo:2021kvv}. Instead, we should use dedicated
(optimal) observables to target one fundamental symmetry at a
time~\cite{Hankele:2006ma,Han:2009ra,Brehmer:2017lrt,Bahl:2021dnc,Barman:2021yfh,Hall:2022bme}.
In the Higgs-gauge sector, the optimal observable is the azimuthal
angle between the two tagging jets in weak boson fusion. For
associated top--Higgs production, the azimuthal angle between a
charged lepton from one top decay and the down quark from the other
plays a similar role.  Accurately extracting it faces the challenge of
identifying the corresponding decay jet. Another powerful observable
probing the Higgs-top interaction is the Collins--Soper
angle~\cite{Collins:1977iv,Goncalves:2018agy,Goncalves:2021dcu,Barman:2021yfh}.
Again, the challenge is to map it onto the observed final state after
particle decays, parton shower, and detector effects.

Both of these observables illustrate the common problem that an
optimal or ideal kinematic correlation is usually not defined on the
reconstructed final state. So while an optimal observable provides
full sensitivity without the need to consider additional phase space
correlations, we pay a prize in its reconstruction.

The standard inference approach for such kinematic correlations is to
approximate them at the reconstruction level. For this approximation,
we can use a directly observable correlation at the reconstruction
level or rely on some kind of algorithm. The approximation is
unlikely to be optimal. An improved approach would be to encode the
observable in a learned mapping, for instance, through neural
networks.  Fundamentally different and, in principle, optimal
alternatives are simulation-based
inference~\cite{Brehmer:2018eca,Brehmer:2019xox} or the matrix element
method~\cite{Kondo:1988yd,Artoisenet:2010cn,Bury:2020ewi,Butter:2022vkj},
but they come at a significant numerical cost and are hard to re-interpret
for other measurements.

For cases where an optimal observable is defined in some kinematic
frame, we propose a simplified unfolding approach, where we unfold the
reconstruction-level events to the appropriate reference frame, and
then construct the optimal observable for the down-stream task.
Unfolding or reconstructing events beyond the immediately available
detector output is a long-standing
problem~\cite{Lucy:1974yx,Zech:2012ch,Spano:2013nca,Brenner:2019lmf},
undergoing transformative progress through modern machine learning
(ML)~\cite{Gagunashvili:2010zw,Glazov:2017vni,Datta:2018mwd,Andreassen:2019cjw,
Bellagente:2019uyp,Bellagente:2020piv,Leigh:2022lpn,Backes:2022vmn,Shmakov:2023kjj}.
One key observation is that forward and backward simulations are
completely symmetric when we interpret them as sampling from
conditional probabilities~\cite{DAgostini:1994fjx,Cowan:2002in}.  This
motivates ML-unfolding through probabilistic inverse
simulations~\cite{Bellagente:2019uyp,Bellagente:2020piv,Backes:2022vmn},
which allows us to reconstruct observables or parameters defined at
any level of our forward simulations, for instance, unfolding detector
effects, parton shower, particle decays~\cite{Leigh:2022lpn}, all the
way to measuring fundamental parameters~\cite{Bieringer:2020tnw}.

This generative unfolding technique allows us to just reconstruct key
observables, which have the advantage that they can be used in the
standard analysis frameworks of ATLAS and CMS, but with a performance
increase from the full set of kinematic correlations learned through
the unfolding. To guarantee stable network predictions and to be able
to quantitatively extract the training-induced network uncertainties,
we use the Bayesian version~\cite{bnn_early3} of the conditional
normalizing flows~\cite{Bellagente:2021yyh,Butter:2021csz}, for which
the likelihood losses should lead to well-calibrated
results. Eventually, this kind of analysis can serve as a simple
starting point for ML-unfolding, as it can be expanded through
additional observables step by step.

In this paper, we use $CP$-violation through a complex top Yukawa
coupling to show how ML-unfolding techniques can construct and
numerically encode observables in the reference frame where they are
defined. In Sec.~\ref{sec:obs}, we first describe our neural network
architecture, the physics task, and the treatment of phase space.  In
Sec.~\ref{sec:tth}, we introduce our reference process and discuss our
results and potential generalization errors.  Finally,
Sec.~\ref{sec:summary} is reserved for summary and outlook.

\section{Reconstructing observables by unfolding}
\label{sec:obs}

In this study, we propose to use statistical unfolding through inverse
simulation~\cite{Bellagente:2019uyp,Bellagente:2020piv} to construct
kinematic observables in a specific partonic reference frame. While we
are making use of unfolding techniques in constructing a given
observable, the precision, control, and model dependence of the
unfolding is not a limiting factor for our analysis. Instead, we treat
the so-defined observable like any other kinematics construction.

\subsection{Generative unfolding}

Generative unfolding is based on the observation that a forward
simulation from a parton-level event $x_\text{part}$ to a reco-level
event $x_\text{reco}$ just samples from an encoded conditional
probability,
\begin{align}
  r \sim \mathcal{N}(r) 
  \; \stackrel{x_\text{part}}{\longrightarrow} \;
  x_\text{reco} \sim p(x_\text{reco}|x_\text{part})
  \qqquad \text{(forward)} \; .
\label{eq:forward}
\end{align}
This simulation can be trivially inverted on the same training data,
so we can unfold detector effects, initial-state jet radiation, or
particle decays, by sampling from the inverse conditional probability,
\begin{align}
  r \sim \mathcal{N}(r) 
  \; \stackrel{x_\text{reco}}{\longrightarrow} \;
  x_\text{part} \sim p(x_\text{part}|x_\text{reco})
  \qqquad \text{(inverse)} \; .
  \label{eq:inverse}
\end{align}
In both cases, the standard training relies on paired events $\{
x_\text{part}, x_\text{reco} \}$. Obviously, this training dataset
leads to model dependence, which can be reduced by using iterative
methods~\cite{Backes:2022vmn}. The target phase space of the inverse
simulation or unfolding can be chosen flexibly, just unfolding
detector effects~\cite{Bellagente:2020piv,Backes:2022vmn}, but also
jet radiation~\cite{Bellagente:2020piv}, particle
decays~\cite{Leigh:2022lpn}, or sampling right into model parameter
space using setups like BayesFlow~\cite{Bieringer:2020tnw}. Inference
through conditional normalizing flows is standard in many fields of
physics~\cite{cinn,Dax:2021tsq}.

Our generative network encoding the conditional probability defined in
Eq.~\eqref{eq:inverse} is a conditional normalizing flow, specifically
a conditional invertible neural network (cINN)~\cite{cinn}, trained
with a likelihood loss to guarantee a statistically correct and
calibrated output.  To link a batch of $B$ phase space points $x_i$ to
a Gaussian latent space $r_i$ with the condition $c_i$, the likelihood
loss reads
\begin{align}
    \loss_\text{cINN} = \sum_{i=1}^B \left( \frac{r_i(x_i;c_i)^2}{2}
    - \log \left| \frac{\partial r_i(x_i;c_i)}{\partial x_i} \right| \right) \; .
\end{align}

\subsection{Periodic splines}

\begin{figure}[b!]
    \centering
    \def\svgwidth{0.5\columnwidth}
    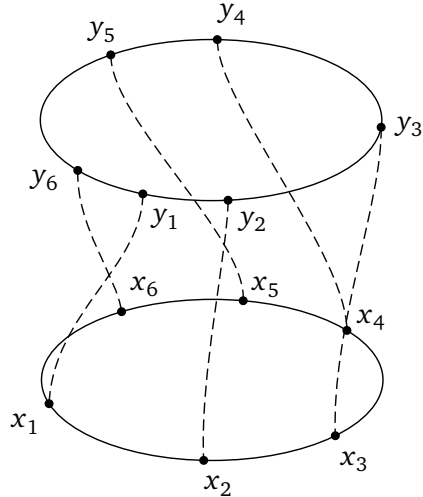
    \caption{Visualization of a modified coupling transformation for
      periodic inputs. This transformation maps $K$ points $x_i$ to
      $K$ points $y_i$ on a circle, while we use rational quadratics
      to interpolate between two points.  The modifications ensure
      that the $x_i$ and $y_i$ can be arbitrarily skewed in relation
      to one another, and that the derivative at $x_1$ is consistent
      with both adjacent rational quadratics.}
    \label{fig:circular_splines_visualization}
\end{figure}

The main part of our cINN is built from coupling layers, specifically
rational quadratic spline blocks~\cite{durkan2019neural}, each
followed by a random permutation. As we will discuss in
Sec.~\ref{sec:obs_ps}, some phase space directions are periodic and
lead to undesired boundary effects when we use these spline
transformations.  To understand this problem in detail, let us
consider a spline transformation
\begin{align}
  g_\theta: \quad [-L, L] \rightarrow [-L, L] \; ,
\end{align}
with parameters $\theta$.  This transformation is given by $K$
different monotonic rational quadratics, parameterized by $K+1$ knot
points $(x_k,y_k)$ and $K+1$ derivatives. The boundaries are $(x_0,
y_0) = (-L, -L)$ and $(x_K, y_K) = (L, L)$.

For periodic inputs the conditions $g(x_0) = y_0 = -L$ and $g(x_K) =
y_K = L$ are unnecessarily restrictive and do not allow the network to
map a distribution onto or past the boundaries, to represent points on
a circle. In addition, we want $g'(x_0) = g'(x_K)$ for periodic
inputs, which is not necessarily true~\cite{Rezende:2020}.  The first
issue can be fixed by replacing $g_\theta$ with
\begin{align}
    \tilde{g}_\theta(x) = g_\theta(x) + g_0 + 2 L k \; ,
\end{align}
with an integer $k$ chosen such that $\tilde{g}(x)$ always lies within
$[-L, L]$, and a new parameter $g_0$ added to $\theta$.  To solve the
second issue, we simply remove one of the derivative parameters from
$\theta$ and set $\tilde{g}_\theta'(x_K) = \tilde{g}_\theta'(x_0)$.
The resulting transformation $\tilde{g}_\theta$ is visualized in
Fig.~\ref{fig:circular_splines_visualization}.

With these modifications, the number of parameters encoding the
transformation does not change.  This means that we can use the same
sub-network to determine $\theta$ for, both, periodic and non-periodic
inputs.  In practice, we split the input vector to the transformation
into periodic and non-periodic inputs and apply $g_\theta$ and
$\tilde{g}_\theta$ separately to each part.  This also implies that we
have to keep track of the permutations between coupling blocks, to be
able to determine the type (periodic or non-periodic) of each input
throughout the network.  As a last detail, we use a uniform latent
space distribution instead of a Gaussian for periodic dimensions.

Bayesian neural networks allow us to efficiently control the training
stability and estimate training-related uncertainties. They extend
standard architectures for regression, classification, or generative
networks to distributions for each network weight, providing a key
tool for explainable AI for instance in fundamental physics
applications. The uncertainty on the output can then be extracted
through
sampling~\cite{bnn_early3,deep_errors,Bollweg:2019skg,Kasieczka:2020vlh,Butter:2022xyj}. For
generative networks, the uncertainty can be defined for the underlying
density
estimation~\cite{Bellagente:2021yyh,Butter:2021csz,Butter:2022vkj,Plehn:2022ftl},
for which the network learns an uncertainty map over the target phase
space. The critical aspect of Bayesian networks is how to make them
numerically viable and still retain all their promising features. We
use a variational approximation for the training, combined with
independent Gaussians for each network weight. Such Bayesian networks
include an optimal regularization, so they can outperform their
deterministic counterparts with limited extra numerical effort. As
always, we emphasize that the underlying approximations do not have to
limit the expressivity of the networks when it comes to the sampled
uncertainties. Moreover, we can treat the formal bias in the Gaussian
widths as a hyperparameter, which needs to be adjusted and should be
checked for stability.

We use the standard Bayesian version of the cINN, as introduced in
Ref.~\cite{Butter:2021csz}, but with periodic splines. The network is
implemented in \pytorch~\cite{pytorch}. In addition, we use the
\adam~\cite{adam} optimizer with a constant learning rate. The
hyper-parameters employed in our study are provided in
Tab.~\ref{tab:cinn}.

\begin{table}[t!]
\centering
\begin{small}
\begin{tabular}{l c}
\toprule
    Parameter & Value \\
\midrule
Block type & periodic rational quadratic spline blocks \\
Number of bins & $10$ \\
Block Period & $2\pi$ \\
Block Domain (non-Periodic) & $[-5.0, 5.0] \rightarrow [-5.0, 5.0]$ \\
\midrule
Number of Blocks & 16 \\
Layers per Block & 5 \\
Units per Layer & 256 \\
Weight Prior Type & Gaussian \\
Weight Prior $\log(\sigma^2)$ & 1.0 \\
\midrule
Number of Epochs (Bayesian) & 100 (200) \\
Batch Size & 1024 \\
Optimizer & \textsc{Adam} \\
Learning Rate & $2.0 \times 10^{-4}$ \\
\midrule
Total number of training events & $\sim$1.2M \\
Training/Testing split & $80\%/20\%$ \\
\bottomrule
\end{tabular}
\end{small}
\caption{Setup and hyper-parameters of the Unfolding-cINN.}
\label{tab:cinn}
\end{table}

\subsection{Phase space parametrization}
\label{sec:obs_ps}

The unfolding method introduced above is identical to full,
high-dimensional unfolding to the parton-level. However, in this
application, we will only target a small number of kinematic
distributions. Moreover, the unfolding network will then be used to
define these distributions as part of the standard LHC analysis
chain. This application allows us to improve the description of
relevant phase space directions at the potential expense of
correlations which are not useful for the measurement. In our case, we
will guarantee the correct descriptions of the intermediate top-mass
peaks through the network architecture.

The simplest way of encoding LHC events at the parton-level is through
the components of the final-state 4-momenta. However, the
corresponding redundant degrees of freedom are not adapted to the
production of intermediate on-shell particles and it's reduced phase
space.  One way to improve the performance of generative networks is
to add a maximum mean discrepancy (MMD) between a given set of
generated and truth distributions~\cite{Butter:2019cae} in the loss
function.  Its main advantage is that it only affects the target
distribution and avoids an unnecessarily large model dependence. The
disadvantage is that the additional loss term complicates the training
and consequently limits the precision of the network.  For our INN
architecture, the computation of an MMD loss requires samples
generated from the latent distribution, while the usual INN loss works
on latent-space samples.

In our case, where the dominant signal and background processes share
intermediate mass peaks, we can learn these features directly, through
an appropriate phase space parametrization. For top decays with 9
degrees of freedom in the final state, a natural parametrization
starts with the corresponding top 4-momentum, and then adds the
invariant $W$-mass and a set of less sensitive angular observables,
\begin{align}
   \left\{ \; m_t, p_{T,t}, \eta_t, \phi_t,
        m_W, \eta^t_W, \phi^t_W,
        \eta^W_{\ell,u}, \phi^W_{\ell,u} \; \right\} \; .
\label{eq:ps_top}
\end{align}
Here $m_{t(W)}$ indicates the reconstructed invariant mass of the
corresponding resonance.  The superscripts $t$ and $W$ indicate the
rest frame where the observable is defined, otherwise we use the
laboratory frame. The indices $\ell$ and $u$ indicate the charged
lepton and the up-type quark for leptonic or hadronic $W$-decays.

A network trained on this parametrization will reproduce the invariant
top and $W$-mass distributions, but with drawbacks in the correlations
of the hadronic $W$-decay. To extract $CP$-information, we also want
to give the network access to the most important $CP$-observables,
which we will discuss in detail in Sec.~\ref{sec:obs_cp}. This means
we will include the Collins-Soper angle
$\theta_\textrm{CS}$~\cite{Collins:1977iv,Goncalves:2018agy,Goncalves:2021dcu,Barman:2021yfh}
and the angle between the charged lepton and the down-quark $\Delta
\phi_{\ell d}$.  One such parametrization for the entire $t\bar{t}$
system with 18 degrees of freedom is
\begin{align}
    &\Big\{ \; \vec{p}_{t\bar{t}}, m_{t_\ell}, |\vec{p}^\text{CS}_{t_\ell}|,
        \theta^\text{CS}_{t_\ell}, \phi^\text{CS}_{t_\ell}, m_{t_h}, \notag \\
    &\qquad\sign(\Delta \phi^{t\bar{t}}_{\ell \nu}) \, m_{W_\ell}, |\vec{p}^{t\bar{t}}_\ell|,
        \theta^{t\bar{t}}_\ell, \phi^{t\bar{t}}_\ell, |\vec{p}^{t\bar{t}}_\nu|, \notag \\
    &\qquad\sign(\Delta \phi^{t\bar{t}}_{du}) \, m_{W_h}, |\vec{p}^{t\bar{t}}_d|,
        \theta^{t\bar{t}}_d, \Delta\phi^{t\bar{t}}_{\ell d}, |\vec{p}^{t \bar{t}}_u| \; \Big\} \; .
\label{eq:ps_pairs}
\end{align}
The superscripts $\text{CS}$ and $t\bar{t}$ indicate the Collins-Soper
frame of the $t\bar{t}$-system and the $t\bar{t}$ rest frame; the
latter rotated such that $\vec{p}^{t\bar{t}}_{t_\ell}$ points in the
direction of the positive $z$-axis.  Also, $t_\ell$ and $t_h$ denote
the leptonically and hadronically decaying tops, while $u$ and $d$
denote the up- and down-quarks from the $W$-decay.  Using
$\sign(\Delta\phi^{t\bar{t}}_{AB}) m_W$ as a phase space direction
makes it harder for the network to generate the $W$-peaks, but solves
the problem of quadratic phase space constraints.

We emphasize that the combination of generative unfolding with the
phase space para\-metrization of Eq.~\eqref{eq:ps_pairs} is expected
to introduce a bias in the unfolding. However, for our application, we
can ignore this bias given our choice of signal channel and our choice
of target observable.  Moreover, a potential bias will render the
network-defined observable sub-optimal, but does not affect its
evaluation in a standard analysis.

\section{CP-phase from Higgs-top production}
\label{sec:tth}

The example we choose to illustrate unfolding as a way to define
dedicated observables is associated Higgs and top quark pair
production
\begin{align}
  pp
  \to t \bar{t} h + \text{jets}
  \to (b u \bar{d}) \; (\bar{b} \ell^- \bar{\nu}) \; (\gamma \gamma) + \text{jets} \; ,
  \label{eq:proc}
\end{align}
plus the charge-conjugated process. $CP$-violating BSM effects
modifying the top Yukawa coupling can be parametrized through the
Lagrangian~\cite{Demartin:2014fia}
\begin{align}
  \lag \supset
  - \frac{m_{t}}{v} \kappa_t \bar{t}(\cos\alpha + i \gamma_5 \sin\alpha) t h\; ,
  \label{eq:Lagrangian}
\end{align}
where $\alpha$ is the $CP$-violating phase, $\kappa_t$ the absolute
value of the top Yukawa coupling, and $v=246$~GeV the Higgs VEV. The
SM-limit is $\kappa_t=1$ and $\alpha=0$.  Deviations from the SM will
affect Higgs production and the decay. While changes in the scalar
Higgs decay will only impact the total rate, we focus on kinematic
effects in the production.

The Lagrangian in Eq.~\eqref{eq:Lagrangian} can be linked to the
standard SMEFT framework used for general LHC analyses at mass
dimension six.  In this case, we introduce two Wilson coefficients to
modify the top Yukawa~\cite{Whisnant:1997qu,Yang:1997iv}
\begin{align}
  \lag
  &\supset
  \frac{f_t}{\Lambda^2} \ope_t
  + \frac{\tilde{f}_t}{\Lambda^2} \tilde{\ope}_t  \notag \\
  &\equiv
  \left(\phi^\dag \phi - \frac{v^2}{2} \right)
  \left(\frac{f_t}{\Lambda^2}\left(\bar{q}_Lt_R\tilde{\phi}+\tilde{\phi}^\dagger\bar{t}_Rq_L\right) 
  +i\frac{\tilde{f}_t}{\Lambda^2}\left(\bar{q}_L t_R\tilde{\phi}-\tilde{\phi}^\dagger\bar{t}_R q_L\right)\right)\;,  
  \label{eq:d6}
\end{align}
where $\phi$ is the Higgs doublet, $\tilde{\phi}= i\sigma_2\phi^*$,
and $q_L$ the heavy quark doublet $(t_L,b_L)$.  The parameters
$\kappa_t$ and $\alpha$ in Eq.~\eqref{eq:Lagrangian} can be computed
as
\begin{align}
  \kappa_t^2 = \left( -1 + \frac{v^3f_t}{\sqrt{2}m_t\Lambda^2} \right)^2
  + \left( \frac{v^3\tilde{f}_t }{\sqrt{2}m_t\Lambda^2} \right)^2
  \qquad \text{and} \qquad 
  \tan \alpha = \dfrac{\tilde{f}_t}{f_t-\dfrac{\sqrt{2}m_t\Lambda^2}{v^3}} \; .
  \label{eq:relate}
\end{align}
We emphasize that the SMEFT description implicitly assumes that new
physics enters through higher-dimensional operators. In contrast, a
$CP$-phase of the top Yukawa can already arise as a dimension-4
modification of the SM-Lagrangian, reflected by the scale combination
$v^3/(m_t \Lambda^2)$ appearing above.

\subsection{CP-observables}
\label{sec:obs_cp}

There are, fundamentally, two ways of testing the $CP$-structure of
the Higgs Yukawa coupling introduced in Eq.~\eqref{eq:Lagrangian}: we
can measure the angle $\alpha$ and conclude from a significant
deviation $\alpha \ne 0$ that $CP$ is violated in the top Yukawa
coupling, ideally using simulation-based
inference~\cite{Brehmer:2016nyr,Brehmer:2019xox,Barman:2021yfh,Bahl:2021dnc}
or the matrix element method~\cite{Butter:2022vkj}. Alternatively, we
can define an optimal observable for $CP$-violation and test the
actual symmetry~\cite{Han:2009ra,Brehmer:2017lrt,Goncalves:2018agy}.

\subsubsection*{Classical reconstruction}

To search for $CP$-violation, spin correlations between the top and
anti-top quarks in $t\bar{t}h$ production are ideal, because the short
top-lifetime allows for a transfer of the top-polarization to the
decay products prior to hadronization or spin
decorrelation~\cite{Bigi:1986jk}. The angular correlation between the
top-spin and the momenta of the top decay products is given by
\begin{align}
  \frac{1}{\Gamma_t}\frac{d\Gamma}{d\cos\xi_i}
  =\frac{1}{2}\left( 1+ \beta_i P_t\cos\xi_i \right)\; ,
\end{align}
where $\xi_i$ is the angle between the top spin and the $i$-th
particle in the top quark rest frame, $P_t \in [0,1]$ is the
polarization of the top quark, and $\beta_i$ is the spin analyzing
power of the $i$-th decay product. Due to the left-handed nature of
the weak interaction, the charged lepton and $d$-quark display the
largest spin analyzing power,
\begin{align}
 \beta_{\ell^+}
 = \beta_{\bar{d}}
 =1 
 \qquad \text{(to leading order)}.
\end{align}
While one cannot tag a $d$-jet, it is possible to find efficient
proxies.  A practical solution is to select the softer of the two
light-flavor jets in the top rest frame.  This choice gives a spin
analyzing power for this jet as 50\% of that of the charged
lepton~\cite{Jezabek:1994qs,Barman:2021yfh,Dong:2023xiw}. Assuming
that the softer $W$-decay jet in the top rest frame comes from the
$d$-quark, we can now construct appropriate angular correlations to
measure.

\subsubsection*{Linear CP-observables}

The basis of optimal observables testing a symmetry are $U$-even or
$U$-odd observables, defined through their transformation properties
on the incoming and outgoing states,
\begin{align}
\obs \left( U\ket{i} \to U\ket{f} \right) 
= \pm \obs \left( \ket{i} \to \ket{f} \right) \;,
\end{align}
where in our case $U = CP$. Furthermore, a genuine $U$-odd observable
is defined as an observable which vanishes in a $U$-symmetric theory
\begin{align}
\Langle \obs \Rangle_{\lag = U \lag U^{-1}} = 0 \; .
\end{align}
The two definitions are related in that any $U$-odd observable is also
a genuine $U$-odd observable under the condition that the initial
state and the phase space are
$U$-symmetric~\cite{Valencia:1994zi,Brehmer:2017lrt}, so the genuine
$U$-odd property is weaker.

Unfortunately, we cannot infer a $CP$-invariant theory from $\langle
\obs \rangle$ of a $CP$-odd observable alone. While a $\langle \obs
\rangle \ne 0$ always points to a $CP$-violating theory, the result
$\langle \obs \rangle = 0$ can appear in $CP$-symmetric and in
$CP$-violating theories.  To further analyze this case, we can
construct a $CP$-odd observable that is also odd under the so-called
naive time reversal $\hat{T}$. Now, the expectation value of this
observable is completely tied to the $CP$-symmetry of the underlying
theory~\cite{Brehmer:2017lrt}.

$CP$-odd observables can be constructed either as $\hat{T}$-even
scalar products of two 4-momenta or as a $\hat{T}$-odd contraction of
four independent 4-momenta through the Levi-Civita tensor.  For the $t
\bar{t}$-system of $t\bar{t}h$ production we can use two top momenta
and decay momenta
\begin{align}
\left\{ p_{b_l}, p_l, p_\nu, p_{b_h}, p_u, p_d \right\} \; .
\label{eq:top_dec}
\end{align}
It is straightforward to construct the $C$-even, $P$-odd, and
$\hat{T}$-odd observable
\begin{align}
\obs = \varepsilon_{\mu\nu\,\sigma\rho} p_{t_h}^\mu p_{t_\ell}^\nu p_A^\rho p_B^\sigma \; ,
\label{eq:genuine_cp_odd_obs}
\end{align}
with suitable top decay momenta $p_{A,B}$.  We can use the
$CP$-invariance of the initial state and the phase space for $t
\bar{t} h$ production to show that its expectation value in
Eq.~\eqref{eq:genuine_cp_odd_obs} vanishes in the SM.

In the $t\bar{t}$ center of mass frame, we can turn
Eq.~\eqref{eq:genuine_cp_odd_obs} into a triple product, a standard
form for $CP$-odd observables,
\begin{align}
  \obs 
    &= 2 \, E_{t_h} \, \vec{p}_{t_\ell} \cdot  (\vec{p}_A \times \vec{p}_B) \; .
\label{eq:genuine_cp_odd_obs_triple_product}
\end{align}
However, it depends on the top 4-momentum, which are hard to determine
accurately.  It can be modified by introducing the azimuthal angle
difference $\Delta \phi^{t\bar{t}}_{AB} = \phi_A^{t\bar{t}} -
\phi_B^{t\bar{t}}$ in the $t\bar{t}$
frame~\cite{Goncalves:2018agy,Barman:2021yfh},
\begin{align}
    \Delta \phi_{AB}^{t\bar{t}} = \textrm{sgn}[\vec{p}_{t_{\ell}}
            \cdot(\vec{p}_A \times \vec{p}_{B})]
        \arccos\left[ \frac{\vec{p}_{t_{\ell}}\times \vec{p}_{A}}%
                {|\vec{p}_{t_{\ell}}\times \vec{p}_{A}|}
            \cdot \frac{\vec{p}_{t_{\ell}}\times \vec{p}_{B}}%
                {|\vec{p}_{t_{\ell}}\times \vec{p}_{B}|}\right] \; ,
\end{align}
to give us
\begin{align}
  \obs = 2p_t^z \, E_t \, p_{T,\,A} \, p_{T,\,B}  \, \sin \Delta \phi^{t\bar{t}}_{AB} \; ,
\label{eq:genuine_cp_odd_obs_modified_triple_product}
\end{align}
where we choose $p_{t_{\ell}} = \{E_t, 0, 0, p_t^z\}$ and $p_{t_h} =
\{E_t, 0, 0, -p_{t}^z\}$.  By construction, $\obs$ and $\Delta
\phi^{t\bar{t}}_{AB}$ are sensitive to the linear interference terms
in the scattering cross section, and therefore sensitive to the sign
of the $CP$-phase.

\begin{figure}[t]
    \centering
    \includegraphics[width=0.5\textwidth]{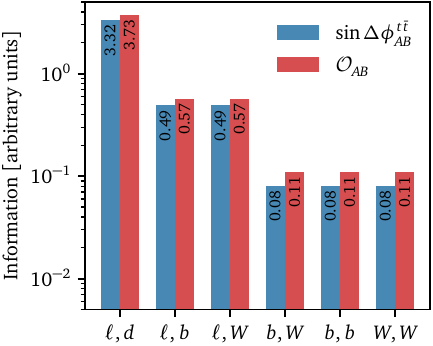}
    \caption{Fisher information $I$ for the linear CP-observables
      $\sin\Delta \phi_{AB}^{t\bar{t}}$ (blue) and $\mathcal{O}_{AB}$
      (red), probing the sensitivity to $CP$-violating phase $\alpha$
      in $t\bar{t}h$ production.}
        \label{fig:information_in_CP_observables}
\end{figure}

These linear $CP$-observables can be constructed for various $\{A,B\}$
pairs, and their $CP$-sensitivity dependents on the spin-analyzing
power of the particles $A$ and $B$. We compute the Fisher information
metric $I$ to rank their $CP$-sensitivity, using
\texttt{MadMiner}~\cite{Brehmer:2019xox,Barman:2021yfh}.  The
$\alpha$-dependent component of $I$ is defined as
\begin{align}
    I = \mathbb{E} \left[
\frac{\partial \log p(x|\kappa_t, \alpha)}{\partial \alpha}
\frac{\partial \log p(x|\kappa_t, \alpha)}{\partial  \alpha}
\right],
\end{align}
where $p(x|\kappa_t,\alpha)$ represents the likelihood of a phase
space configuration $x$ given the theory parameters $\kappa_t$ and
$\alpha$.  $\mathbb{E}$ denotes the expectation value at the SM point,
$(\kappa_t,\alpha)_\text{SM}=(1,0)$. In
Fig.~\ref{fig:information_in_CP_observables}, we show the Fisher
information at parton-level associated with the linear
$CP$-observables $\obs_{AB}$ in red and the Fisher information for
$\sin\Delta \phi_{AB}^{t\bar{t}}$ in blue.
 
First, we see that for all combinations $(A,B)$ the Fisher information
in $\obs_{AB}$ is slightly larger than the Fisher information in $\sin
\Delta \phi_{AB}^{t\bar{t}}$, an effect of the momentum-dependent prefactor in
$\obs_{AB}$.  Among the various combinations $(A,B)$, the combination
of the lepton and the down-type quark is the most sensitive. This
corresponds to the maximal spin analyzing power for this pair. Next
comes the combination where either the charged lepton or the down
quark is replaced by the $b$-quark or the $W$-boson. In this case, the
Fisher information is suppressed by two powers of
\begin{align}
 \beta_b = \beta_W \sim 0.4 \; . 
\end{align}    
The correlation between a pair of $b$-quarks or $W$-bosons is further
suppressed by another factor $\beta_{b,W}^2$.

\subsubsection*{Non-linear observables and Collins-Soper angle}
\label{sec:obs_more}

For a given realization of $CP$-violation in an SM-like interaction
vertex, the $CP$-observable defined in the previous section is not
guaranteed to be the most powerful
observable~\cite{Englert:2012xt}. This is obvious for dimension-6
operators, where the symmetry structure is often combined with a
momentum dependence of the interaction~\cite{Brehmer:2017lrt}, and the
two aspects can, in principle, be tested independently.  Comparing the
two handles, $CP$-odd observables are only sensitive to the
interference between the SM-contribution and the $CP$-violating matrix
element, while observables testing the momentum structure of the
interaction vertex can be dominated by the new-physics-squared
contribution. For large $CP$-phases $\alpha$, the more promising
analysis strategy will use a general test of the structure of the
top-Higgs coupling. This motivates using a combination of dedicated
$CP$-observables with general interaction probes as an optimal search
strategy.

Several observables have been evaluated as probes of the $CP$-phase
$\alpha$ in Eq.~\eqref{eq:Lagrangian} using $t\bar{t}h$
production~\cite{Goncalves:2018agy,Goncalves:2021dcu,Barman:2021yfh}.
They include the pseudorapidity difference between the two tops and
the azimuthal angle between the two tops, the Higgs transverse
momentum~\cite{Demartin:2014fia, Demartin:2015uha}, or the invariant
mass of the top and anti-top pair,
\begin{align}
\left\{ \; 
\Delta\eta_{t\bar{t}}, 
\Delta \phi_{t\bar{t}},  
p_{T,h}, 
m_{t\bar{t}}
\; \right\} \,.
\end{align}
These standard observables can be supplemented with the projection
angle~\cite{Demartin:2014fia,Demartin:2015uha,Gunion:1996xu}
\begin{align}
    b_4 = 
        \frac{p_{z,t}}{|\vec{p}_t|} \cdot 
        \frac{p_{z,\bar{t}}}{|\vec{p}_{\bar{t}}|} \; .
\end{align}

Finally, we can use the Collins-Soper angle
$\theta_\text{CS}$~\cite{Collins:1977iv}, the angle between the top
quark and the bisector of the incoming hadrons in the $t\bar{t}$
center of mass frame.  The original motivation for the Collins-Soper
angle was to define an observable for the Drell-Yan process $pp \to
\ell^+ \ell^-$ that corresponds to the scattering angle.
Factorization arguments suggest the di-lepton rest frame, to minimize
ISR-effects and then study the angular correlations between the
incoming quarks and the outgoing leptons.  In this frame the 3-momenta
of the quarks and leptons each define a plane, and in turn an
azimuthal angle and a polar angle between the two planes.

Without ISR the $z$-axis of the so-defined CS-frame is trivially given
by the parton and hadron directions. Including ISR, we instead define
this $z$-axis as halving the angle between one of the hadrons and the
reverse direction of the other hadron. The Collins-Soper angle can be
used to measure the polarization of the intermediate gauge boson, the
weak mixing angle~\cite{CMS:2011utm}, or the (Lorentz) structure of
the interaction vertices.  The Collins-Soper angle can also be used to
probe the structure of the Higgs-photon
coupling~\cite{ATLAS:2013xga,CMS:2014afl} and to boost new physics
searches in $h^*\to ZZ$, $Zh$, and $t\bar{t}Z$
channels~\cite{Goncalves:2018fvn,Goncalves:2018ptp,Goncalves:2020vyn,MammenAbraham:2022yxp}.
Finally, it can be generalized to $t\bar{t}$ production, where it is
constructed for the top momentum in the $t\bar{t}$ rest
frame~\cite{Frederix:2007gi,Bernreuther:2015fts,Goncalves:2018agy}.
While the Collins-Soper angle has no specific sensitivity to
$CP$-violation, we view it as the Swiss Army knife of coupling tests.

\begin{figure}[t!]
    \includegraphics[page=1]{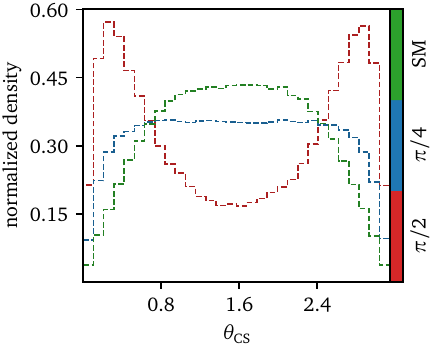}\hfill%
    \includegraphics[page=2]{cpobsv_lhc.pdf}\par%
    \vspace*{0.01\columnwidth}
    \includegraphics[page=3]{cpobsv_lhc.pdf}\hfill%
    \includegraphics[page=4]{cpobsv_lhc.pdf}\par%
    \caption{Distributions for the CS-angle $\theta_\text{CS}$, $\Delta 
    \eta_{t_{\ell}t_{h}}$, $b_4$, and $\Delta\phi_{t_\ell t_h}$, all at parton-level 
        for semileptonic $t\bar{t}h$ production. 
        The distributions are shown for SM~(green), $\alpha = \pi/4$~(blue) and
        $\alpha=\pi/2$~(red).}   
    \label{fig:parton_level_cs_deta_b4}
\end{figure}

All above-mentioned kinematic observables are sensitive to the
new-physics-squared terms, proportional to $\sin^2 \alpha$ or
$\cos^2\alpha$, in the $t\bar{t}h$ rate, with no sensitivity to the
sign of the CP-phase.  From Ref.~\cite{Barman:2021yfh}, we know the
relative sensitivity of these observables to probe the Higgs-top
$CP$-structure through a modified Fisher information metric,
accounting for non-linear effects. The top-five observables with the
highest Fisher information for $\alpha$ are (symbolically written)
\begin{align}
\Delta\eta_{t\bar{t}}
>\theta_\text{CS}
> b_4
> \Delta \phi_{t\bar{t}}
> p_{T,h} \; .
\label{eq:obs}
\end{align}
We show the parton-level distributions for the four most sensitive
observables in the semileptonic $t\bar{t}h$ channel for the SM value
$\alpha = 0$ and $\alpha = \pi/4, \pi/2$ at the LHC with
$\sqrt{s}=14$~TeV in Fig.~\ref{fig:parton_level_cs_deta_b4}.
Different values of $\alpha$ lead to distinctly different profiles in
the distributions.

As alluded to above, the technical challenge and a limitation to the
optimality of a given analysis is to construct the different
observables in their respective kinematic frames.  Considering their
strong sensitivity on $\alpha$, we include the leading observables in
the phase space parametrization given in Eq.~\eqref{eq:ps_pairs} to
target this problem directly.

\subsection{Unfolding-based analysis}
\label{sec:unfold}

The standard challenge for every LHC analysis is to extract a small
signal from a large (continuum) background.  For our simple study, we
show how we can avoid modeling this step.  The generative unfolding
trained on $t\bar{t}h$ events gives us the probability $p(x_\text{part}|x_\text{reco},S)$ that
a parton-level signal event $x_\text{part}$, corresponds to an assumed signal
event $x_\text{reco}$ at reconstruction level.  What we are ultimately interested
in, however, is a model parameter $\alpha$, which could be a mass, a
$CP$-phase, or any other continuous theory parameter, which affects
our signal distribution.  Since we do not know if a particular
reco-level event $x_\text{reco}$ is signal or background, we only care about the
full probability $p(\alpha | x_\text{reco})$ of our model parameter, given some
reco-level event $x_\text{reco}$ which is either signal or background.  Since
$\alpha$ does not change the background, this probability can be split
into the distribution $p(\alpha | x_\text{part})$, where $x_\text{part}$ is a parton-level
signal event, and the probability $p(x_\text{part} | x_\text{reco})$ of $x_\text{part}$ given $x_\text{reco}$:
\begin{align}
    p(\alpha | x_\text{reco}) = \int p(\alpha | x_\text{part}) p(x_\text{part} | x_\text{reco}) \dif{x}.
\end{align}
The challenge is to compute $p(x_\text{part} | x_\text{reco})$ from our unfolding result $p(x_\text{part}
| x_\text{reco}, S)$.  Using the definition of conditional probabilities we can
write
\begin{align}
    p(x_\text{part} | x_\text{reco}) &= \sum_{T \in \{S, B\}} p(x_\text{part} | x_\text{reco}, T) p(T | x_\text{reco}) \notag \\
        &= p(x_\text{part} | x_\text{reco}, S) p(S | x_\text{reco}) + p(x_\text{part}|x_\text{reco},B) (1 - p(S | x_\text{reco})) \; ,
\end{align}
where the probabilities of $x_\text{reco}$ being a signal or background event,
$p(T|x_\text{reco})$, can be encoded in a trained classifier.  Let us consider for
a moment what the probability $p(x_\text{part} | x_\text{reco}, B)$ tells us.  We are
interested in signal events $x_\text{part}$, i.e. events that are affected by
$\alpha$.  By definition, background events $x_\text{reco}$ cannot give us any
information, beyond prior knowledge, about $x_\text{part}$.  For this reason, we
can drop $x_\text{reco}$ and write $p(x_\text{part} | x_\text{reco}, B) = p(x_\text{part})$, where $p(x_\text{part})$ is only
constrained through prior knowledge.  This includes our model
assumptions as well as phase-space constraints due to a finite
center-of-mass energy.  We can now write
\begin{align}
    p(x_\text{part} | x_\text{reco}) = p(x_\text{part} | x_\text{reco}, S) p(S | x_\text{reco}) + p(x_\text{part}) (1 - p(S | x_\text{reco})).
    \label{eq:background_handling}
\end{align}
What Eq.~\eqref{eq:background_handling} shows, is that we can limit
our unfolding model to extracting $p(x_\text{part} | x_\text{reco}, S)$ and still
include background events into our analysis later, without changing
our model.

As given in Eq.~\eqref{eq:proc}, we study $pp \to t_h \bar{t}_\ell h$
production with $h \to \gamma\gamma$ at the HL-LHC. The dominant
background is continuum $t\bar{t}\gamma\gamma$ production, subdominant
contributions arise from the process $Wb\bar{b}(h \to \gamma\gamma)$.
We use \texttt{MadGraph5\_aMC@NLO}~\cite{Alwall:2014hca} with
\texttt{NNPDF2.3QED}~\cite{Ball:2013hta} to generate signal events at
leading order with $\sqrt{s}=14~$TeV. Signal events are simulated
without kinematic cuts using the $\textrm{HC\_NLO\_X0}$ UFO
model~\cite{Artoisenet:2013puc,deAquino:2013uba}.  Parton showering
and hadronization effects are simulated using
\texttt{Pythia\,8}~\cite{Sjostrand:2007gs}. The detector response is
simulated with \texttt{Delphes\,3}~\cite{deFavereau:2013fsa}, using
the default ATLAS HL-LHC card~\cite{HLLHC_card,Cepeda:2019klc}.

Next, we select events containing exactly two photons, two $b$-tagged
jets, one lepton, and at least two light-flavored jets. The individual
particles in the final state are required to satisfy the acceptance
cuts
\begin{alignat}{9}
  p_{T,b} & > 25~\gev\;,
  &\quad
  p_{T,j} & > 25~\gev\;,
  &\quad
  p_{T,\ell} & > 15~\gev\;,
  &\quad
  p_{T,\gamma} & > 15~\gev\;,
  &\notag \\
  |\eta_{b}| & < 4\;,
  &\quad
  |\eta_{j}| & < 5\;,
  &\quad
  |\eta_{\ell}| & < 4\;,
  &\quad
  |\eta_{\gamma}| & < 4 \;.
  \label{eqn:selection_cuts}
 \end{alignat}
At the parton-level, the signal phase space involves eight final state
particles; following Sec.~\ref{sec:obs_ps} it requires 22 parameters
if we are assuming the Higgs is fully and uniquely reconstructed. The
training dataset involves an event-wise pairing of parton and detector
level events with up to six light-flavored jets, satisfying the
selection cuts in Eq.~\eqref{eqn:selection_cuts}.  While the event at
the reconstruction level requires additional degrees of freedom for
jet radiation, the number of degrees of freedom is reduced by the
neutrino. An additional challenge is the combinatorics of the $b$-jets
and light-flavor jets.

\subsection{Results}
\label{sec:tth_ref}

\begin{figure}[b!]
    \includegraphics[page=1]{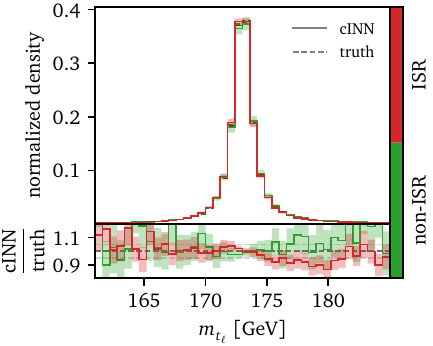}\hfill%
    \includegraphics[page=2]{isr_jetnum_dependence.pdf}\par%
    \vspace*{0.01\columnwidth}
    \includegraphics[page=3]{isr_jetnum_dependence.pdf}\hfill%
    \includegraphics[page=4]{isr_jetnum_dependence.pdf}\par%
    \caption{Jet combinatorics --- cINN-generated distributions for $m_{t_{\ell}}$, $m_{t_h}$,
        $m_{ht_h}$ and $m_{t_{\ell}t_{h}}$ in the SM. Unfolded distributions are shown as solid
        lines, parton-level truth as dashed lines. The training data set either does not
        include ISR (green) or up to six ISR jets (red).}
    \label{fig:minv_tl_th_tt}
\end{figure}

\paragraph{Jet combinatorics} 
The first results from unfolding $t\bar{t}h$ SM-events are presented
in Fig.~\ref{fig:minv_tl_th_tt}. We train the unfolding network on
SM-events and also apply it to SM-events.  First, we examine the
robustness of the network to unfold a variable number of jets to the
parton-level.  For our lepton-hadron reference process in
Eq.~\eqref{eq:proc} two light-flavor jets come from the hadronic top
decay, while additional jets arise from QCD jet radiation.  The
unfolding network has to reconstruct the two hard jets at the parton
level from a variable number of jets at the detector
level~\cite{Bellagente:2020piv}.

To evaluate the unfolding performance, we examine four invariant
masses: $m_{t_{\ell}}$, $m_{t_h}$, $m_{ht_{h}}$, and
$m_{t_{\ell}t_{h}}$. We train one network on SM events without ISR and
one network on events with up to six light-flavored jets.  The
corresponding cINN-generated distributions are shown as solid lines in
Fig.~\ref{fig:minv_tl_th_tt}. The parton-level truth is displayed as
dashed lines. We find that unfolded distributions generated by both
networks are in good agreement with the parton-level truth in the bulk
of the phase space. Despite the added combinatorial ambiguity, the
performance of both networks is largely comparable. We also show the
uncertainties from the Bayesian setup, represented as $1\sigma$ error
bands. They test the stability of the unfolding network similar to an
ensemble of networks. It is important to observe that the truth
distributions remain within these error bands.

\begin{figure}[t]
    \includegraphics[page=1]{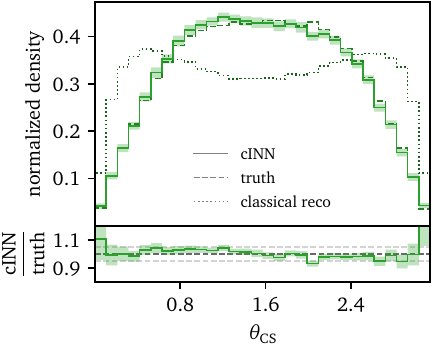}\hfill%
    \includegraphics[page=2]{model_precision_sm.pdf}\par%
    \vspace*{0.01\columnwidth}
    \includegraphics[page=3]{model_precision_sm.pdf}\hfill%
    \includegraphics[page=4]{model_precision_sm.pdf}\par%
    \caption{Reconstructing dedicated observables --- cINN-generated
      distributions and distributions based on classical
      reconstruction~\cite{Barman:2021yfh} for $\theta_\text{CS}$,
      $\Delta\eta_{t_\ell t_h}$, $b_4$ and $\Delta\phi_{t_\ell t_h}$
      for SM events. The secondary panels show the bin-wise agreement
      between the cINN-generated distributions and the parton-level
      truth.}
    \label{fig:model_precision_sm_main}
\end{figure}

\paragraph{Reconstructing dedicated observables} 
For Fig.~\ref{fig:model_precision_sm_main} we train the unfolding
network on SM events with up to six light-flavor jets. We compare
cINN-generated events at the parton-level and in the appropriate rest
frame with events from a classical reconstruction for four
particularly interesting observables from Sec.~\ref{sec:obs_cp}:
$\theta_\textrm{CS}$, $\Delta \eta_{t\bar{t}}$, $b_{4}$, and $\Delta
\phi_{t\bar{t}}$.  For comparison, we display the parton-level truth
as dashed lines.  In the ratio we observe that the generated
distributions agree with the truth within a few percent.  Slightly
larger deviations in the tails are due to limited training statistics.

\begin{figure}[p]
    \includegraphics[page=1]{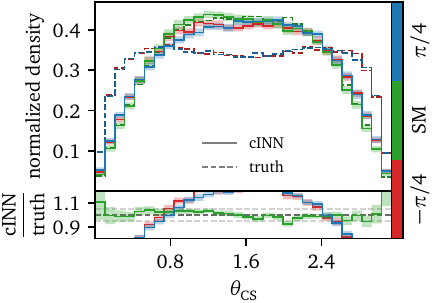}\hfill%
    \includegraphics[page=2]{unfold_sm_full.pdf}\par%
    \vspace*{0.01\columnwidth}
    \includegraphics[page=3]{unfold_sm_full.pdf}\hfill%
    \includegraphics[page=4]{unfold_sm_full.pdf}\par%
    \vspace*{0.01\columnwidth}
    \includegraphics[page=1]{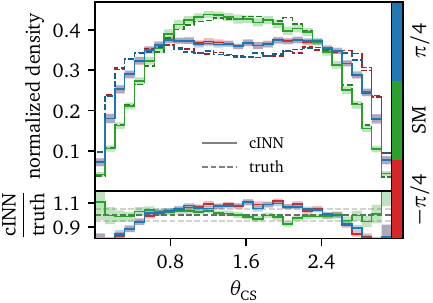}\hfill%
    \includegraphics[page=2]{unfold_hypotheses_under_sm.pdf}\par%
    \vspace*{0.01\columnwidth}
    \includegraphics[page=3]{unfold_hypotheses_under_sm.pdf}\hfill%
    \includegraphics[page=4]{unfold_hypotheses_under_sm.pdf}\par%
    \caption{Model dependence --- cINN-generated distributions for $\theta_{\text{CS}}$, 
    $\Delta\eta_{t_\ell t_h}$, $b_4$, and $\Delta\phi_{t_\ell t_h}$.
    Upper two rows: unfolding of SM events using three
    different networks, trained on data with $\alpha = -\pi/4, 0, \pi/4$. 
    Lower two rows: unfolding of events with $\alpha = -\pi/4, 0, \pi/4$, with a 
    network trained on SM events.}
    \label{fig:unfold_sm_full_with_detector_reco}
\end{figure}

The conventional approach to complex kinematic correlations in the
semileptonic $t\bar{t}$ system relies on a complex reconstruction
algorithm, with a significant loss of information due to missing
correlations~\cite{Barman:2021yfh}. We show the reconstructed
distributions from the classical reconstruction strategy developed in
Ref.~\cite{Barman:2021yfh} as dotted lines. Comparing these
distributions to the cINN-unfolded version, we see that at least for a
network trained and tested on SM events, the improvement from
generative unfolding is striking.

\begin{figure}[t]
    \includegraphics[page=1]{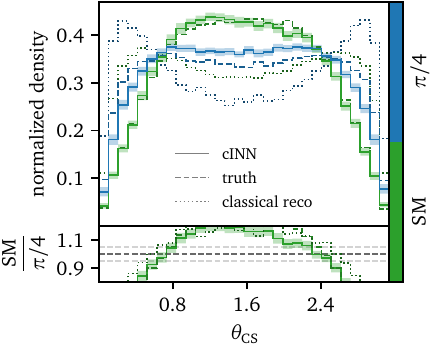}\hfill%
    \includegraphics{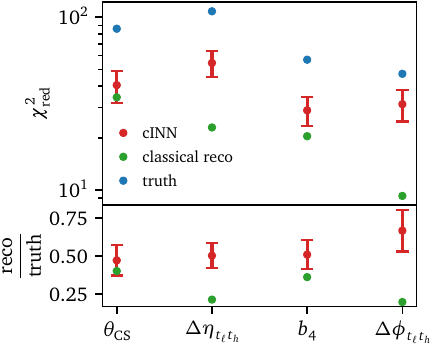}\par%
    \caption{Sensitivity --- Left: cINN-generated distributions for $\theta_\text{CS}$
        from unfolding events with two $\alpha$ values. These generated distributions are
        compared to the distributions obtained from classical reconstruction methods, as
        described in Ref.~\cite{Barman:2021yfh}, and the respective truth.
        Right: To quantify the sensitivity of the cINN, as shown here for $\theta_\text{CS}$,
        we compute the reduced $\chi^2$-value between the distributions ($\sim$120k events and
        64 bins) for both $\alpha$ values, using the Poisson errors of the bin counts.
        We do this for the cINN-generated (red), classically reconstructed (green), and truth
        distributions (blue) of $\theta_\text{CS}$, $\Delta\eta_{t_\ell t_h}$, $b_4$, and
        $\Delta\phi_{t_\ell t_h}$.
        In the bottom panel, we also show the ratio of the reduced $\chi^2$-values of
        the cINN and the classical reconstruction to the truth.
        Uncertainties on cINN-generated values are obtained from the Bayesian setup.}
    \label{fig:compare_ml_to_classical_reco}
\end{figure}

\paragraph{Model dependence}
After observing the significant improvement through our new method for
SM events, we need to test how model-dependent the network training
is.
In the upper panels of Fig.~\ref{fig:unfold_sm_full_with_detector_reco},
we reconstruct the usual set of key observables for SM events, but with three
different networks, trained on events generated with the $CP$-angles $\alpha =
-\pi/4, 0, \pi/4$.
We adopt the BSM values $\alpha = \pm \pi/4$ here, as these choices closely align
with the current experimental limits~\cite{atlas2020, cms2020}.
From Fig.~\ref{fig:unfold_sm_full_with_detector_reco} we expect, for
instance, the network trained on events with $\alpha= \pi/4$ to be
biased towards a broader $\theta_\text{CS}$ distribution, a wider
rapidity difference $\Delta \eta_{t_\ell,t_h}$, and a flatter $b_4$
distribution. In the different panels we see a slight bias, especially
in the secondary panels. But the bias remains at the order of 10\%, at
most 20\%, much below the change in the corresponding distributions
from varying $\alpha$. On the other hand, this bias is significantly
larger than the uncertainty band, which indicates that this model
dependence can be reduced through the proposed iterative method of
Ref.~\cite{Backes:2022vmn}. The corresponding study is beyond the
scope of this paper, because it balances a reduced bias of the
unfolding with less statistics, an aspect which we do not include in
this proof-of-principle study.

In the lower panels of
Fig.~\ref{fig:unfold_sm_full_with_detector_reco} we test the model
dependence the other way around, by unfolding data for different
$\alpha = -\pi/4, 0, \pi/4$ using a network trained on SM events. The
figure of merit are the ratios of cINN-unfolded and respective truth
distributions, shown in the secondary panels. This situation is closer
to the reality of a measurement, where we infer $\alpha$ by comparing
the distribution extracted from data to different simulated
hypotheses. As before, we see a slight bias, now towards the SM
structure of a more narrow $\theta_\text{CS}$ distribution, a narrow
rapidity difference $\Delta \eta_{t_\ell,t_h}$, and a steeper $b_4$
distribution. Also, as before, the effect of the bias is much smaller
than the effect of $\alpha$ on the data, leaving us optimistic that we
can use the cINN-unfolded distribution to measure $\alpha$.

\paragraph{Sensitivity}
Finally, in Fig.~\ref{fig:compare_ml_to_classical_reco}, we apply the generative unfolding to
SM and $\alpha = \pi/4$ events.
The unfolding network is trained on SM events.
As a baseline comparison, we also show the same two curves for classical reconstruction of
$\theta_\text{CS}$, following Ref.~\cite{Barman:2021yfh} as dotted lines in the left panel of
Fig.~\ref{fig:compare_ml_to_classical_reco}.
As mentioned earlier, generative unfolding leads to a major improvement over classical
reconstruction.
The difference in the two unfolded kinematic distributions, shown in solid lines, illustrates
the reach of an analysis based on the kinematic distribution.
To showcase the improvement in new physics sensitivity, we calculate the reduced $\chi^2$
values for $\theta_\text{CS}$, $\Delta \eta_{t_{\ell}t_h}$, $b_4$, and
$\Delta \Phi_{t_{\ell}t_h}$ between the SM and $\alpha = \pi/4$ hypotheses, using the Poisson
errors of the bin counts.
The reduced $\chi^2$ values are computed with $\sim$120k events and 64 bins, for three
scenarios: parton-level truth (blue), classical reconstruction from Ref.~\cite{Barman:2021yfh}
(green), and the cINN-based generative model trained on SM events (red).
A higher $\chi^{2}$ value indicates a greater sensitivity to new physics.

The results show that the unfolding setup leads to an enhancement in sensitivity compared to
the classical reconstruction strategy.
This indicates that the generative unfolding approach is effective in extracting more
information from the kinematic distributions, thereby improving the analysis' capability to
detect and explore new physics phenomena.
We further observe that the network is slightly more consistent in reproducing the sensitivity
relations of the true observable distributions than the classical reconstruction.
The latter performs well on some observables, but quite bad for others.
Especially surprising is the classical sensitivity on $\theta_\text{CS}$, given that the
reconstruction here is far from the actual CS-angle.

\section{Outlook}
\label{sec:summary}

Unfolding is one of the most exciting development of analysis
preservation and publication at the LHC. Modern machine learning makes
it possible to unfold high-dimensional distributions, covering all
correlations without binning. Generative unfolding defines this
unfolding in a statistically consistent manner. However, using
unfolded data is a challenge for the ATLAS and CMS analysis chains,
especially in controlling and estimating uncertainties. 

We investigated a simpler application of the unfolding technique, the
extraction of a kinematic observable in a specific partonic reference
frame. It solves the dilemma that on the one hand an optimal
observable requires no complex correlations, but on the other hand
such an observable is, typically, hard to reconstruct. In this case
the generated kinematic distribution can be used like any other
observable; the unfolding network is nothing but a kinematic
reconstruction algorithm.

The perfect examples for a challenging kinematic correlation are the
Collins-Soper angle or the optimal $CP$-observables in $t\bar{t}h$
production. They allow us to measure a $CP$-phase in the top Yukawa
coupling, a cosmologically relevant parameter entering an LHC
signature at dimension four and at leading order. We have shown that
unfolding allows us to extract the leading observables for such a
$CP$-phase $\alpha$, with the help of an appropriate phase space
parametrization. While such a parametrization might shape the unfolded
kinematic distribution, this effect can be controlled through
calibration.

First, we have shown that the cINN-unfolding can solve the
combinatorics of $W$-decay jets vs QCD jet radiation. Second, the
unfolded distributions of SM events, with a network trained on SM
events, show excellent agreement with the parton-level
truth. Potential differences are covered by the uncertainty estimate
from the Bayesian network. Third, we have tested the model dependence
in two different ways --- unfolding SM event using networks trained on
events with different amounts of $CP$-violation and unfolding events
with $CP$-violation using a network trained on SM events. For the
former, we have found that there exists a small, but significant model
dependence, which can be removed through Bayesian iterative
improvements. For the latter, the unfolded distributions do not
perfectly reproduce the respective truth, but the bias is much smaller
than the kinematic effect of the $CP$-angle.

All these tests have motivated a comparison of the reach of the HL-LHC
for the $CP$-angle $\alpha$, based on classical reconstruction methods
and on cINN-unfolded distributions. The generative unfolding approach 
effectively extracts more information from kinematic distributions, 
enhancing sensitivity to new physics phenomena. This highlights the 
importance of advanced machine learning techniques, such as cINNs, 
for the HL-LHC.

While this study is clearly not the last word on this analysis
technique, we consider the outcome promising enough for an
experimental study, with a proper treatment of statistical
limitations, continuum backgrounds, calibration, and iterative
improvements of the unfolding network.

\section*{Acknowledgements}

RKB and DG thank the U.S. Department of Energy for financial support, under grant number
DE-SC0016013.
Some computing for this project was performed at the High Performance Computing Center at
Oklahoma State University, supported in part through the National Science Foundation grant
OAC-1531128.
TH is funded by the Carl-Zeiss-Stiftung through the project \textsl{Model-Based AI: Physical
Models and Deep Learning for Imaging and Cancer Treatment}.
This research is supported by the Deutsche Forschungsgemeinschaft (DFG, German Research
Foundation) under grant 396021762 -- TRR~257: \textsl{Particle Physics Phenomenology after the
Higgs Discovery} and through Germany's Excellence Strategy EXC~2181/1 -- 390900948 (the
\textsl{Heidelberg STRUCTURES Excellence Cluster}).

\bibliographystyle{hepml}
\bibliography{literature}
\end{document}